# CBCT-Based Synthetic CT Image Generation Using Conditional Denoising Diffusion Probabilistic Model

Junbo Peng[1,2], Richard L.J. Qiu[1], Jacob F Wynne[1], Chih-Wei Chang[1], Shaoyan Pan[1], Tonghe Wang[3], Justin Roper[1], Tian Liu[4], Pretesh R. Patel[1], David S. Yu[1] and Xiaofeng Yang[1,2,*]

[1]Department of Radiation Oncology and Winship Cancer Institute, Emory University, Atlanta, GA 30322, USA

[2]Nuclear and Radiological Engineering and Medical physics Programs, George W. Woodruff School of Mechanical Engineering, Georgia Institute of Technology, Atlanta, GA 30332, USA

[3]Department of Medical Physics, Memorial Sloan Kettering Cancer Center, New York, NY 10065, USA

[4]Department of Radiation Oncology, Icahn School of Medicine at Mount Sinai, New York, NY 10065, USA

*Email: xiaofeng.yang@emory.edu

# Abstract

**Background:** Daily or weekly cone-beam computed tomography (CBCT) scans are commonly used for accurate patient positioning during the image-guided radiotherapy (IGRT) process, making it an ideal option for adaptive radiotherapy (ART) replanning. However, the presence of severe artifacts and inaccurate Hounsfield unit (HU) values prevent its use for quantitative applications such as organ segmentation and dose calculation. To enable the clinical practice of online ART, it is crucial to obtain CBCT scans with a quality comparable to that of a CT scan.

**Purpose:** This work aims to develop a conditional diffusion model to perform image translation from the CBCT to the CT domain for the image quality improvement of CBCT.

**Methods:** The proposed method is a conditional denoising diffusion probabilistic model (DDPM) that utilizes a time-embedded U-net architecture with residual and attention blocks to gradually transform standard Gaussian noise to the target CT distribution conditioned on the CBCT. The model was trained on deformed planning CT (dpCT) and CBCT image pairs, and its feasibility was verified in brain patient study and head-and-neck (H&N) patient study. The performance of the proposed algorithm was evaluated using mean absolute error (MAE), peak signal-to-noise ratio (PSNR) and normalized cross-correlation (NCC) metrics on generated synthetic CT (sCT) samples. The proposed method was also compared to four other diffusion model-based sCT generation methods.

**Results:** In the brain patient study, the MAE, PSNR, and NCC of the generated sCT were 25.99 HU, 30.49 dB, and 0.99, respectively, compared to 40.63 HU, 27.87 dB, and 0.98 of the CBCT images. In the H&N patient study, the metrics were 32.56 HU, 27.65 dB, 0.98 and 38.99 HU, 27.00, 0.98 for sCT and CBCT, respectively. Compared to the other four diffusion models, the proposed method showed superior results in both visual quality and quantitative analysis.

**Conclusions:** The proposed conditional DDPM method can generate sCT from CBCT with accurate HU numbers and reduced artifacts, enabling accurate CBCT-based organ segmentation and dose calculation for online ART.

# 1. Introduction

Cone-beam computed tomography (CBCT) scanning is widely used daily or weekly in current image-guided radiotherapy (IGRT) practice for patient setup and treatment monitoring, to decrease positioning error and improve the accuracy of radiotherapy.[1] Compared to traditional diagnostic CT images, CBCT images often suffer from considerable artifacts such as streaking, shading, cupping, and scatter contamination, resulting in severe inaccuracies of the Hounsfield unit (HU) values.[2-4] These factors prevent the use of CBCT images for quantitative applications, hindering the implementation of CBCT-based adaptive radiotherapy (ART). To address this problem, current ART practice uses planning fan-beam CT (pCT) images that have been deformed to match the anatomical structure of CBCT for dose calculation.[5] However, the quality of image registration often depends on the experience and intuition of the operator due to the changes in anatomy and artifacts on CBCT images,[6] limiting the accuracy of ART implementations.

Current attempts to enable direct CBCT-based ART include two major approaches: one is to perform artifacts correction to improve the image quality of CBCT,[7-17] and the other is to generate synthetic CT (sCT) images with the similar level of image quality as pCT from CBCT images.[6,18-22]

There are many approaches to reducing CBCT artifacts, which can be broadly classified as hardware-based correction, model-based, and deep learning-based correction methods. Hardware modifications include anti-scatter grid,[7] primary-modulation beam filter,[8] and lattice-shaped beam stopper.[9] While these hardware modifications show promise in correcting scattering artifacts in CBCT imaging, they can also inherently reduce the system's quantum efficiency and degrade the signal-to-noise ratio of reconstructed images.[20] Model-based approaches attempt to correct artifacts by modeling the CBCT imaging process. For example, Monte Carlo (MC)[10] and analytical[14] algorithms were developed to simulate the scattering signal in CBCT imaging. The MC method is accurate for scatter estimation but is computationally intensive while the analytical scatter kernel algorithm is faster but is less effective for heterogeneous inner structures due to nonlinear scattering process.[23] Deep learning-based methods obtain the mapping from CBCT to artifact-free CBCT by training a neural network using paired data with and without artifacts, which can be performed in both projection and image domains.[15-17] While it is proven to be effective to alleviate scatter,[16,17] streaking,[15] or metal artifacts,[24] the learning-based approach is limited in its ability to suppress specific sources of artifacts and cannot correct other types of artifacts, which is a common issue shared among all artifact-reduction methods.

sCT generation methods employ prior information from the pCT to train a model which maps voxels from source CBCT distribution to target pCT distribution via deep learning.[6,18-22] The most straightforward way is to train a neural network like U-net through supervised learning using matched CBCT-deformed pCT (dpCT) data pairs.[18] However, there are always residual mismatch errors due to anatomic changes between pCT and CBCT images, which can negatively impact the supervised image-to-image translation. To overcome this issue, mainly two categories of methods have been proposed. The first approach is to perform unsupervised learning with unpaired CBCT-pCT data.[19,21] For example, Cycle-GAN[25] enforces cyclic consistency between the source image and the recovered image from the translated image, which can improve the accuracy of anatomical preserving during the CBCT-to-CT translation.[21] The second approach is to introduce a dedicated module and additional loss function to preserve the structure of the input CBCT and learn the structural variations in the supervised learning framework. Gradient difference loss and attention mechanism were introduced to paired Cycle-GAN,[20,22] which combines the feature preserving properties of inherent cyclic loss and introduced strategies, showing superiority in sCT generation in challenging scenarios like pancreatic ART.

Diffusion model is an emerging generative approach that has been shown to outperform GAN in tasks such as image synthesis,[26] image inpainting,[27] and image super resolution.[28,29] It has attracted much attention in the medical imaging field. Diffusion model was first introduced to solve inverse problems like sparse-view CT reconstruction and metal artifact removal.[30] Denoising diffusion probabilistic model (DDPM),[31] one of the most famous diffusion models, has been used for under-sampled MRI reconstruction,[32,33] conversion between MRI and CT images,[34] low-dose CT denoising,[35] and 4D CT generation.[36] Compared to GAN and variational autoencoder (VAE) models which are difficult to interpret and train, diffusion models are analytically principled, easy to train, and produce state-of-the-art (SoAT) image quality.[37,38]

In this work, we proposed a conditional DDPM framework for sCT generation from CBCT images. Paired CBCT-dpCT data were used for model training where the dpCT served as the target data distribution and the CBCT served as the condition. The trained model can then gradually translate a standard Gaussian noise to the target sCT, conditioned on the CBCT counterpart. Our experiments on brain and head and neck (H&N) patients showed that the proposed conditional DDPM can generate sCT with accurate HU values and reduced artifacts, which makes direct CBCT-based ART implementation possible.

## 2. Materials and Methods
### 2.1. Image acquisition and preprocessing

In this study, CBCT and pCT images were collected from 41 brain and 47 H&N patients. For the clinical brain cases, a total of 4682 slices from 30 patients were used for training and 500 slices randomly selected from the remaining 11 patients were used for testing. For the H&N patient study, 4314 slices from 37 patients were employed for training and 500 slices from other 10 patients were randomly selected for testing. All CT scans were acquired on Siemens SOMATOM Definition AS with 120 kVp, and CBCT scans were acquired on Varian TrueBeam with 100 kVp. The voxel sizes were $1.0 \times 1.0 \times 1.0$ mm$^3$ for both CT and CBCT images, respectively. CT images were deformably registered and resampled to match their corresponding CBCT images using Velocity AI 3.2.1, so that the data volume and voxel size between the CBCT and CT pairs were identical. Body contours extracted from CBCT and CT images were merged, and then applied to both images. All images were cropped to the size of $256 \times 256$ and normalized to [-1, 1] before being fed into the network.

### 2.2. Denoising diffusion probabilistic models (DDPM)

DDPM is a certain parameterization of diffusion models, which is a class of latent variable models using a Markov chain to convert a standard Gaussian distribution to the target data distribution.[31] Suppose the target data

$x_0 \sim q(x_0)$. As the orange-arrow flow shown in figure 1(a), a sequence of gradually corrupted images $x_1, x_2, \cdots, x_T$ can be constructed after each Markov forward diffusion process formulated as:

$$q(x_t|x_{t-1}) = \mathcal{N}(x_t; \sqrt{1-\beta_t}x_{t-1}, \beta_t I) \tag{1}$$

$$q(x_{1:T}|x_0) = \prod_{t=1}^{T} q(x_t|x_{t-1}) \tag{2}$$

where $T$ is the total number of diffusion steps, $\beta_t \in (0,1)$ is a hyper-parameter controlling the variance of incremental Gaussian noise, and $\mathcal{N}(x_t; \mu, \sigma)$ represents a Gaussian distribution of mean $\mu$ and variance $\sigma$. According to the properties of the Gaussian distribution, $x_t$ at any arbitrary step can be computed directly conditioned on the initial image $x_0$ using the reparameterization strategy of $\alpha_t := 1 - \beta_t$ and $\bar{\alpha}_t := \prod_{i=1}^{t} \alpha_i$:

$$q(x_t|x_0) = \mathcal{N}(x_t; \sqrt{\bar{\alpha}_t}x_0, (1-\bar{\alpha}_t)I) \tag{3}$$

$$x_t = \sqrt{\bar{\alpha}_t}x_0 + \sqrt{1-\bar{\alpha}_t}\epsilon \tag{4}$$

where $\epsilon \sim \mathcal{N}(0,I)$ and $x_T$ becomes an isotropic Gaussian distribution when $T \to \infty$. The simplified formula in equation (4) can be used for one-step calculation of $x_t$ in the training stage introduced below.

Based on the Bayes theorem, the posterior of each step in the reverse process is also a Gaussian distribution conditioned on $x_t$ and $x_0$:

$$q(x_{t-1}|x_t, x_0) = \frac{q(x_t|x_{t-1})q(x_{t-1}|x_0)}{q(x_t|x_0)}$$

$$= \mathcal{N}(x_{t-1}; \tilde{\mu}_t(x_t, x_0), \tilde{\beta}_t I) \tag{5}$$

with $\tilde{\mu}_t(x_t, x_0) = \frac{\sqrt{\bar{\alpha}_{t-1}}\beta_t}{1-\bar{\alpha}_t}x_0 + \frac{\sqrt{\bar{\alpha}_t}(1-\bar{\alpha}_{t-1})}{1-\bar{\alpha}_t}x_t$ and $\tilde{\beta}_t = \frac{1-\bar{\alpha}_{t-1}}{1-\bar{\alpha}_t}\beta_t$ (6)

From an isotropic Gaussian distribution $q(x_T)$ given a large enough $T$, one can gradually generate a sample in the target distribution $q(x_0)$ based on the posterior distribution $q(x_{t-1}|x_t)$. However, $q(x_{t-1}|x_t)$ is not computable for the unknown distribution of $x_0$. The DDPM framework tries to predict the $\tilde{\mu}_t(x_t, x_0)$ from $x_t$ through a network with parameters of $\theta$ and $t$:

$$\tilde{\mu}_t(x_t, x_0) \approx \mu_{\theta,t}(x_t) \tag{7}$$

then the equation (5) can be approximated as

$$q(x_{t-1}|x_t, x_0) \approx \mathcal{N}(x_{t-1}; \mu_{\theta,t}(x_t), \tilde{\beta}_t I) \tag{8}$$

Equations (4) and (6) show that the prediction of $\tilde{\mu}_t(x_t, x_0)$ is equivalent to predict the noise $\epsilon$ from $x_t$, which can be formulated as

$$\epsilon_{\theta,t}(x_t) \approx \epsilon \tag{9}$$

then the loss function could be written as

$$Loss = \|\epsilon - \epsilon_{\theta,t}(x_t)\|_2^2 \tag{10}$$

With the trained network to approximate $\epsilon$ at each step, one can reverse the diffusion process to recover an original image located in the distribution $q(x_0)$ from a Gaussian noise $x_T$, as the blue-arrow flow in figure 1(a):

$$p_\theta(x_{t-1}|x_t) = \mathcal{N}(x_{t-1}; \frac{1}{\sqrt{\alpha_t}}(x_t - \frac{\beta_t}{\sqrt{1-\bar{\alpha}_t}}\epsilon_{\theta,t}(x_t)), \tilde{\beta}_t I) \tag{11}$$

$$p_\theta(x_{0:T}) = p(x_T) \prod_{t=1}^{T} p_\theta(x_{t-1}|x_t) \qquad (12)$$

As algorithms 1 summarized in Table 1, both the training and sampling processes of DDPM are unconditional. A noise estimator $\epsilon_{\theta,t}(x_t)$ unconditioned on any prior information is trained, and the generation of samples $x_0$ is also guidance-free from any additional conditions.

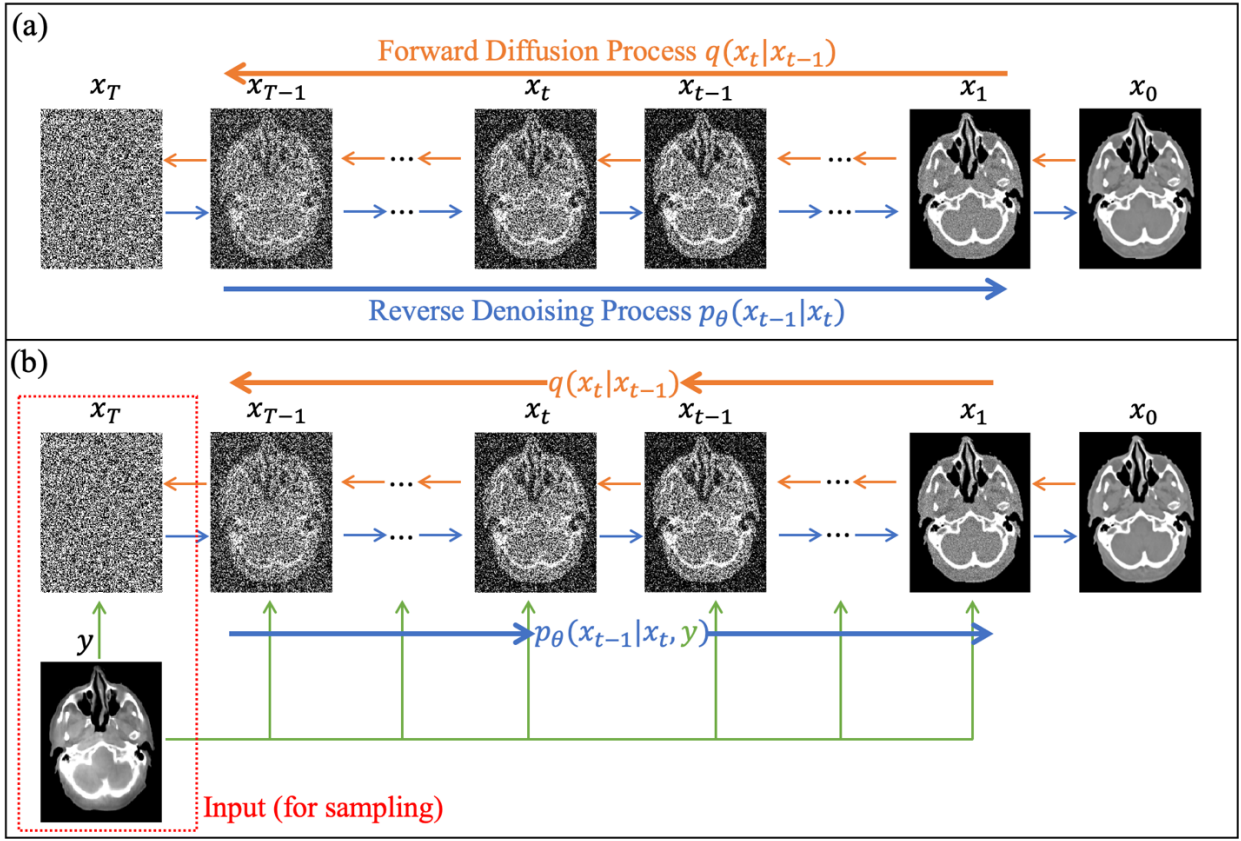

**Figure 1**. Workflows of (a) the conventional DDPM, which consists of unconditional forward and reverse processes, and (b) the proposed conditional DDPM for sCT generation from CBCT, composed of unconditional forward and conditional reverse processes.

### 2.3. The proposed conditional DDPM for synthetic CT from CBCT images

The original DDPM is an unsupervised learning method for unconditional image generation, which is not suitable for generating images with desired semantics. However, CT synthesis from CBCT images is a conditional image-to-image translation task based on CBCT samples. That is, the generated sCT should be the counterpart of input CBCT instead of a random sample in the target CT distribution. Several approaches have been proposed to enforce the condition on diffusion models to control the sample generation,[29,32,33,39] and in this study the input CBCT image $y$ is concatenated with sample $x_t$ along the channel dimension to serve as a static guidance in each sampling step.[29] With the condition $y$ introduced in the reverse process, the noise estimator $\epsilon_{\theta,t}(x_t)$ in DDPM becomes $\epsilon_{\theta,t}(x_t, y)$ and the posterior distribution is $p_\theta(x_{t-1}|x_t, y)$ now. Given the corresponding CT and CBCT pair $(x, y)$, the noise-prediction loss function (10) and reverse Markovian processes (11) and (12) can be modified as

$$Loss = \|\epsilon - \epsilon_{\theta,t}(x_t, y)\|_2^2 \tag{13}$$

$$p_\theta(x_{t-1}|x_t, y) = \mathcal{N}(x_{t-1}; \frac{1}{\sqrt{\alpha_t}}(x_t - \frac{\beta_t}{\sqrt{1-\bar{\alpha}_t}}\epsilon_{\theta,t}(x_t, y)), \tilde{\beta}_t I) \tag{14}$$

$$p_\theta(x_{0:T}|y) = p(x_T)\prod_{t=1}^{T} p_\theta(x_{t-1}|x_t, y) \tag{15}$$

The workflow of the proposed method is shown in figure 1(b), and training and sampling procedures of conditional DDPM are listed as algorithms 2 in table I.

**Table I**. The training and sampling procedures of DDPM (algorithm 1) and the proposed method (algorithm 2).

| **Algorithm 1.1** Training | **Algorithm 1.2** Sampling |
|---|---|
| 1: **repeat** | 1: $x_T \sim \mathcal{N}(0, I)$ |
| 2: $\quad (x_0) \sim p(x)$ | 2: **for** $t = T, \cdots, 1$ **do** |
| 3: $\quad t \sim U([0,1])$ | 3: $\quad z \sim \mathcal{N}(0, I)$ if $t > 1$, else $z = 0$ |
| 4: $\quad \epsilon \sim \mathcal{N}(0, I)$ | 4: $\quad x_{t-1} = \frac{1}{\sqrt{\alpha_t}}\left(x_t - \frac{1-\alpha_t}{\sqrt{1-\bar{\alpha}_t}}\epsilon_{\theta,t}(x_t)\right) + \sqrt{\tilde{\beta}_t}z$ |
| 5: $\quad x_t = x_0 + \sqrt{1-\bar{\alpha}_t}\epsilon$ | 5: **end for** |
| 6: $\quad$ Take a gradient descent step on $\nabla_\theta \|\epsilon_{\theta,t}(x_t) - \epsilon\|^2$ | 6: **return** $x_0$ |
| 7: **until** converged | |
| **Algorithm 2.1** Training | **Algorithm 2.2** Sampling |
| 1: **repeat** | 1: $x_T \sim \mathcal{N}(0, I)$ |
| 2: $\quad (x_0, y_0) \sim p(x, y)$ | 2: **for** $t = T, \cdots, 1$ **do** |
| 3: $\quad t \sim U([0,1])$ | 3: $\quad z \sim \mathcal{N}(0, I)$ if $t > 1$, else $z = 0$ |
| 4: $\quad \epsilon \sim \mathcal{N}(0, I)$ | 4: $\quad x_{t-1} = \frac{1}{\sqrt{\alpha_t}}\left(x_t - \frac{1-\alpha_t}{\sqrt{1-\bar{\alpha}_t}}\epsilon_{\theta,t}(x_t, y)\right) + \sqrt{\tilde{\beta}_t}z$ |
| 5: $\quad x_t = x_0 + \sqrt{1-\bar{\alpha}_t}\epsilon$ | 5: **end for** |
| 6: $\quad$ Take a gradient descent step on $\nabla_\theta \|\epsilon_{\theta,t}(x_t, y) - \epsilon\|^2$ | 6: **return** $x_0$ |
| 7: **until** converged | |

### 2.4. Compared conditional DDPM approaches

In this study, we compared four other strategies for performing conditional DDPM with the proposed method. The first two approaches belong to the category of conditional sampling with an unconditionally trained model.

The proposed method, on the other hand, is a fully conditional model with guided training and sampling. These two methods attempt to control the generation process of unconditional DDPM by guiding each sampling step towards the desired subset.[39] The first approach is called iterative latent variable refinement (ILVR), which is denoted as Uncond-1 strategy in this work. As shown in figure 2(a), an unconditional DDPM is pretrained using CT images to translate the Gaussian noise to the CT image, where $p_\theta(x_{t-1}|x_t)$ the trained posterior distribution unconditioned on CBCT. During the inference stage, ILVR provides condition of CBCT $y$ to unconditional transition $p_\theta(x_{t-1}|x_t)$ without retraining the model. Specifically, the method refines each unconditional transition $x'_{t-1}$ with a reference CBCT $y_{t-1}$ through

$$x_{t-1} = \phi(y_{t-1}) + (I - \phi)x'_{t-1} \tag{16}$$

$$\text{with } x'_{t-1} \sim p_\theta(x'_{t-1}|x_t) \tag{17}$$

where $\phi$ denotes the linear low-pass filter consisting of a sequence of downsampling and upsampling operators. In this way, ILVR ensures that the generated image shares high-level semantics from the condition of given reference image. The second method, denoted as Uncond-2, to refine each step of unconditional inference is weighted summation of $x'_{t-1}$ and $y_{t-1}$ by

$$x_{t-1} = \frac{t-1}{T} y_{t-1} + \frac{T-t+1}{T} x'_{t-1} \tag{18}$$

where the condition of the input CBCT is gradually weakened by decreased weighting factors, as shown in figure 2(b). The pseudo code of sampling process for Uncond-1 and Uncond-2 methods are summarized in table II.

The other two strategies are categorized to the conditionally trained diffusion model, which is similar to the proposed method. However, these two perform adaptively noised condition instead of constant condition on each time step of noise estimation, which are represented by Adap-Cond-1 and Adap-Cond-2 methods. The modified Adap-Cond-1 employs $y_{T-t}$ as condition in $x_{t-1}$ prediction, in which way, stronger conditions are performed on early stages to better guide the sampling process to the desired direction

$$p_\theta(x_{t-1}|x_t, y_{T-t}) = \mathcal{N}(x_{t-1}; \frac{1}{\sqrt{\alpha_t}}(x_t - \frac{\beta_t}{\sqrt{1-\bar{\alpha}_t}}\epsilon_{\theta,t}(x_t, y_{T-t})), \tilde{\beta}_t I) \tag{19}$$

$$\text{with } y_{T-t} \sim \mathcal{N}(y_{T-t}; \sqrt{\bar{\alpha}_t} y_0, (1-\bar{\alpha}_{T-t})I) \tag{20}$$

In Adap-Cond-2 framework, adaptive condition $y_t$ is used to guide the $x_{t-1}$ generation in the reverse Markov process, which can be formulated as

$$p_\theta(x_{t-1}|x_t, y_t) = \mathcal{N}(x_{t-1}; \frac{1}{\sqrt{\alpha_t}}(x_t - \frac{\beta_t}{\sqrt{1-\bar{\alpha}_t}}\epsilon_{\theta,t}(x_t, y_t)), \tilde{\beta}_t I) \tag{21}$$

$$\text{with } y_t \sim \mathcal{N}(y_t; \sqrt{\bar{\alpha}_t} y_0, (1-\bar{\alpha}_t)I) \tag{22}$$

In this way, the additional condition of CBCT and original condition of CT are in the same noise level in the posterior distribution $p_\theta(x_{t-1}|x_t, y_t)$. In summary, Adap-Cond-1 and Adap-Cond-2 replace the constant condition $y$ in the proposed conditional DDPM method with weaker conditions of $y_{T-t}$ and $y_t$, which are shown in figure 3. The pseudo code for two adaptively conditional DDPM schemes are listed in table III.

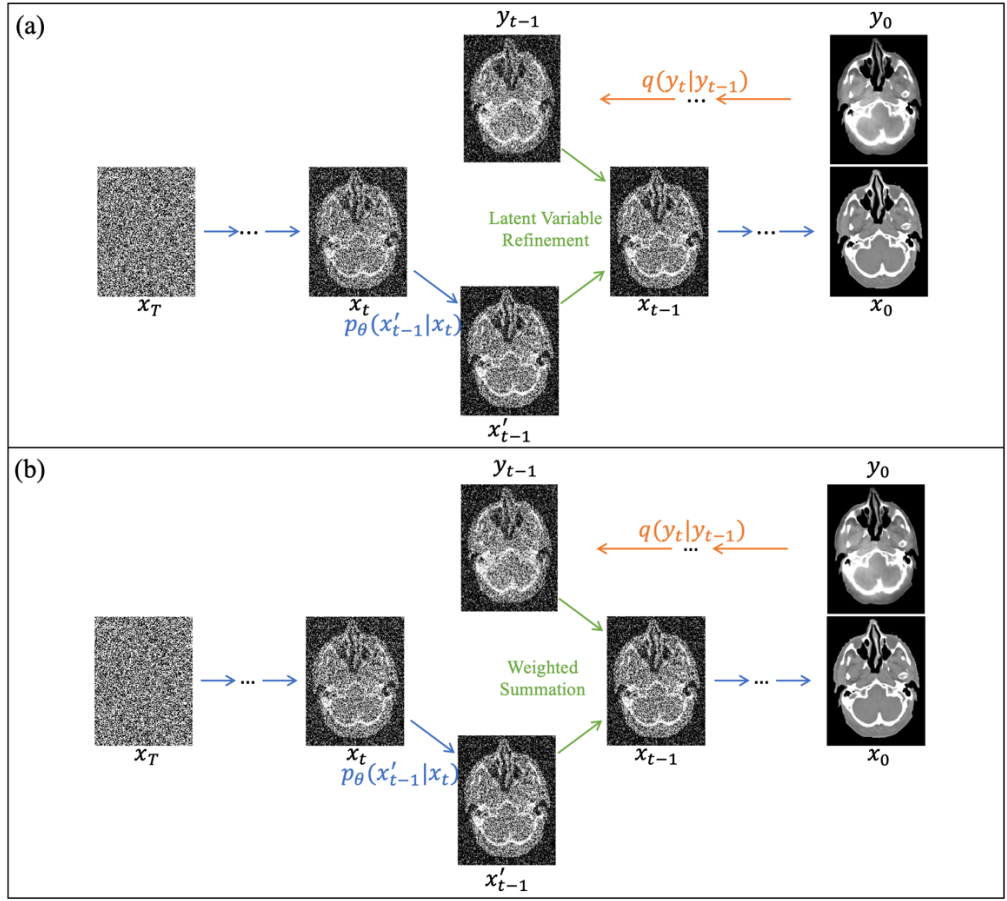

**Figure 2**. The inference stages of (a) Uncond-1 and (b) Uncond-2 methods to perform CT sample generation conditioned on the reference CBCT image.

**Table II**. The sampling procedures of Uncond-1 (algorithms 3) and Uncond-2 (algorithms 4) methods for sCT generation.

| Algorithm 3 | Algorithm 4 |
|---|---|
| 1: $x_T \sim \mathcal{N}(0, \boldsymbol{I})$ | 1: $x_T \sim \mathcal{N}(0, \boldsymbol{I})$ |
| 2: **for** $t = T, \cdots, 1$ **do** | 2: **for** $t = T, \cdots, 1$ **do** |
| 3: $z, \epsilon \sim \mathcal{N}(0, \boldsymbol{I})$ if $t > 1$, else $z = 0$ | 3: $z, \epsilon \sim \mathcal{N}(0, \boldsymbol{I})$ if $t > 1$, else $z = 0$ |
| 4: $x'_{t-1} = \frac{1}{\sqrt{\bar{\alpha}_t}}\left(x_t - \frac{1-\alpha_t}{\sqrt{1-\bar{\alpha}_t}}\epsilon_{\theta,t}(x_t)\right) + \sqrt{\tilde{\beta}_t}z$ | 4: $x'_{t-1} = \frac{1}{\sqrt{\bar{\alpha}_t}}\left(x_t - \frac{1-\alpha_t}{\sqrt{1-\bar{\alpha}_t}}\epsilon_{\theta,t}(x_t)\right) + \sqrt{\tilde{\beta}_t}z$ |
| 5: $y_{t-1} = y_0 + \sqrt{1-\bar{\alpha}_{t-1}}\epsilon$ | 5: $y_{t-1} = y_0 + \sqrt{1-\bar{\alpha}_{t-1}}\epsilon$ |
| 6: $x_{t-1} = \phi(y_{t-1}) + (I - \phi)x'_{t-1}$ | 6: $x_{t-1} = \frac{t-1}{T}y_{t-1} + \frac{T-t+1}{T}x'_{t-1}$ |
| 7: **end for** | 7: **end for** |
| 8: **return** $x_0$ | 8: **return** $x_0$ |

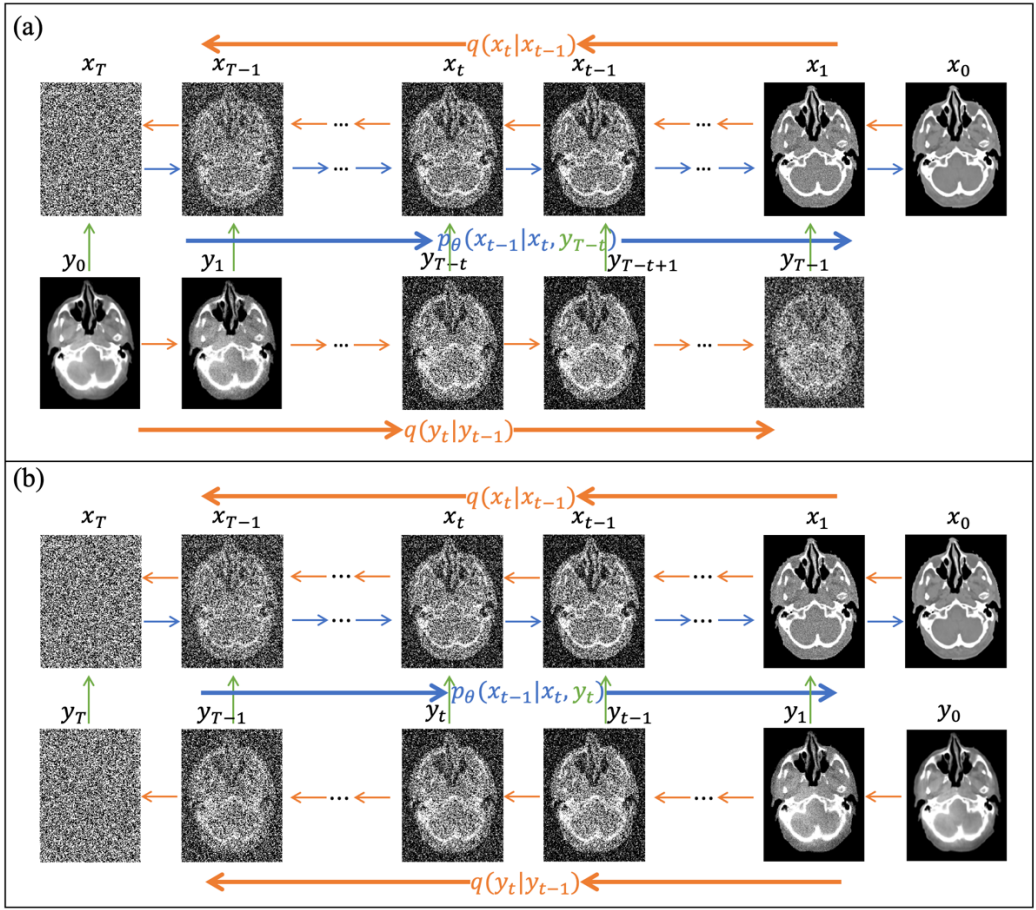

**Figure 3**. The training and sampling processes of (a) Adap-cond-1 and (b) Adap-Cond-2 strategies for sCT generation based on CBCT images.

**Table III**. The training and inference stages of Adap-Cond-1 (algorithms 5) and Adap-Cond -2 (algorithms 6) methods for sCT generation.

| **Algorithm 5.1** Training | **Algorithm 5.2** Sampling |
|---|---|
| 1: **repeat** | 1: $x_T \sim \mathcal{N}(0, \mathbf{I})$ |
| 2: $(x_0) \sim p(x)$ | 2: **for** $t = T, \cdots, 1$ **do** |
| 3: $t \sim U([0,1])$ | 3: $z, \epsilon \sim \mathcal{N}(0, \mathbf{I})$ if $t > 1$, else $z = 0$ |
| 4: $\epsilon, \epsilon' \sim \mathcal{N}(0, \mathbf{I})$ | 4: $y_{T-t} = y_0 + \sqrt{1 - \bar{\alpha}_{T-t}} \epsilon$ |
| 5: $x_t = x_0 + \sqrt{1 - \bar{\alpha}_t} \epsilon$ | 5: $x_{t-1} = \frac{1}{\sqrt{\bar{\alpha}_t}} \left( x_t - \frac{1-\alpha_t}{\sqrt{1-\bar{\alpha}_t}} \epsilon_{\theta,t}(x_t, y_{T-t}) \right) + \sqrt{\tilde{\beta}_t} z$ |
| 6: $y_{T-t} = y_0 + \sqrt{1 - \bar{\alpha}_{T-t}} \epsilon'$ | 6: **end for** |
| 7: Take a gradient descent step on $\nabla_\theta \| \epsilon_{\theta,t}(x_t, y_{T-t}) - \epsilon \|^2$ | 7: **return** $x_0$ |

| Algorithm 6.1 Training | Algorithm 6.2 Sampling |
|---|---|
| 1: **repeat** | 1: $x_T \sim \mathcal{N}(0, I)$ |
| 2: $(x_0, y_0) \sim p(x, y)$ | 2: **for** $t = T, \cdots, 1$ **do** |
| 3: $t \sim U([0,1])$ | 3: $z, \epsilon \sim \mathcal{N}(0, I)$ if $t > 1$, else $z = 0$ |
| 4: $\epsilon, \epsilon' \sim \mathcal{N}(0, I)$ | 4: $y_t = y_0 + \sqrt{1 - \bar{\alpha}_t}\epsilon$ |
| 5: $x_t = x_0 + \sqrt{1 - \bar{\alpha}_t}\epsilon$ | 5: $x_{t-1} = \frac{1}{\sqrt{\bar{\alpha}_t}}\left(x_t - \frac{1-\alpha_t}{\sqrt{1-\bar{\alpha}_t}}\epsilon_{\theta,t}(x_t, y_t)\right) + \sqrt{\tilde{\beta}_t}z$ |
| 6: $y_t = y_0 + \sqrt{1 - \bar{\alpha}_t}\epsilon'$ | 6: **end for** |
| 7: Take a gradient descent step on $\nabla_\theta \|\epsilon_{\theta,t}(x_t, y_t) - \epsilon\|^2$ | 7: **return** $x_0$ |
| 8: **until** converged | |

## 2.5. Implementation and evaluation

Instead of training $T$ totally different networks to predict $\epsilon_{\theta,t}(x_t, y)$ at each step, a single noise-prediction model with time-embedding[31] was used in all $T$ steps. A U-net structure with attention modules and residual blocks was used to predict the noise in each time step.[26] For all three methods of conditional DDPM, the total number of time steps $T$ was set to 1000 and the noise variance was linearly scheduled from $\beta_1 = 10^{-4}$ to $\beta_T = 0.02$. Adam optimizer was used with a learning rate of $10^{-4}$, betas of 0.9 and 0.999, and eps of $10^{-8}$. The batch size was fixed at 2 and drop out ratio was set to 0.3. The linear low-pass filtering operator in ILVR method was bicubic downsampling and upsampling function with a factor of 8.[40] All the experiments were conducted using PyTorch 1.12 on a 24GB Nvidia RTX 3090 GPU. The training was stopped after $1 \times 10^6$ iterations, which took about 50 h, and it took about 2 min per synthetic slice generation.

For quantitative evaluations, we calculated the MAE, PSNR, and NCC between the sCT and CT images, which were taken as the ground truth. They are defined as follows:

$$MAE = \frac{1}{n_x n_y}\sum_{i,j}^{n_x, n_y}|sCT(i,j) - CT(i,j)| \quad (23)$$

$$PSNR = 10 \times \log_{10}\left(\frac{MAX^2}{\frac{1}{n_x n_y}\sum_{i,j}^{n_x, n_y}|sCT(i,j)-CT(i,j)|^2}\right) \quad (24)$$

$$NCC = \frac{1}{n_x n_y}\sum_{i,j}^{n_x, n_y}\frac{(sCT(i,j)-\overline{sCT})(CT(i,j)-\overline{CT})}{\sigma_{sCT}\sigma_{CT}} \quad (25)$$

where $sCT(i,j)$ and $CT(i,j)$ are the value of pixel $(i,j)$ in the sCT and CT respectively. $n_x n_y$ is the total number of pixels. $MAX$ is the maximum pixel value in the sCT and CT images. $\overline{sCT}$ and $\overline{CT}$ are the mean of sCT and CT images. $\sigma_{sCT}$ and $\sigma_{CT}$ are the standard deviation of sCT and CT images. MAE is the magnitude of the voxel-based Hounsfield unit (HU) difference between the original CT and the sCT. PSNR measures if the predicted sCT intensity is evenly or sparsely distributed. NCC is a measure of similarity between CT and sCT as a function of displacement.

## 3. Results
### 3.1. Visual quality improvement

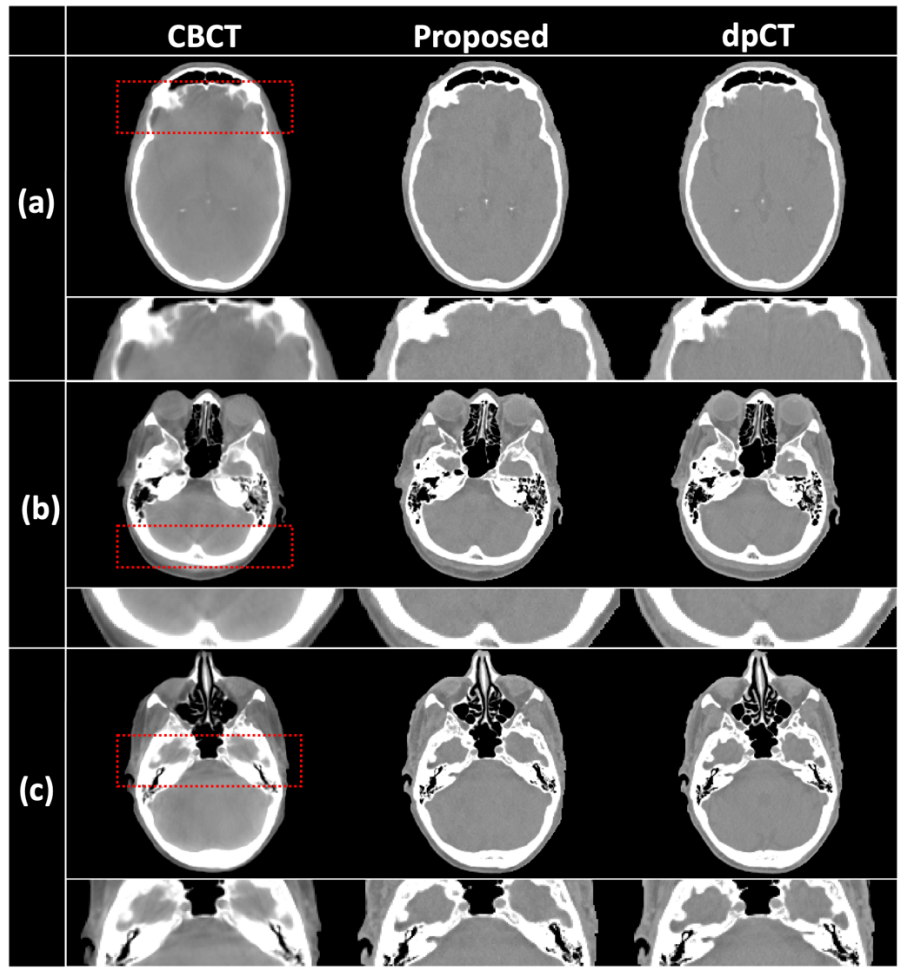

**Figure 4**. Artifacts correction performance in the brain patient study. The first to third columns are the CBCT, sCT generated by the proposed method, and dpCT, respectively. Columns (a) to (c) show different slices from the test dataset, where the zoom-in images of ROIs indicated by red dashed boxes are directly shown below each image. Display window is [-500 500] HU.

The effect of the proposed method on artifacts correction is presented in figure 4 for the brain patient study. Beam hardening artifacts caused by the high-density bone were seen in all three CBCT images, leading to non-uniform area in brain and blurring boundary between bone and soft tissue. However, the artifacts are significantly suppressed on the generated sCT images while preserving the fine structures like contrast agent and maxillary sinus, which can improve the accuracy of organ segmentation. Zoom-in images of the artifacts region are shown below each image for better visualization. Figure 5 shows the visual quality improvement of the proposed sCT in the H&N patient study. CBCT images were severely impacted by streaking artifacts from motion and metal implant, which are indicated by red arrows on each image. For slices (a) and (b), the proposed method generated artifact-free image like dpCT counterparts. Moreover, for the extremely degraded images by metal artifact like (c) and (d), the proposed sCT images achieved better artifact-suppressing performance than the dpCT images. HU histogram plots of all 500 testing slices for each study are summarized in Figure 6. Compared to the CBCT curves, the shape and peak for sCT curves are much closer to dpCT curves, showing improved HU fidelity of sCT from CBCT corresponding to dpCT.

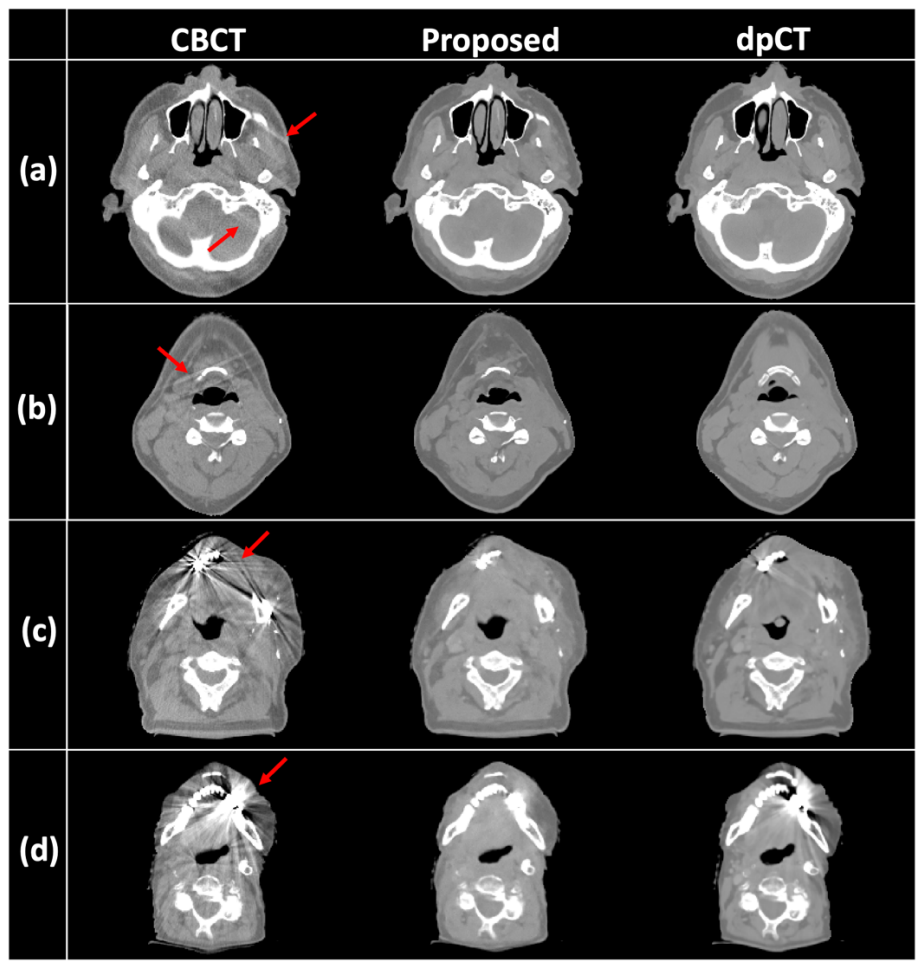

**Figure 5**. Artifacts correction performance in the H&N patient study. Red arrows indicate artifacts on CBCT images. Display window is [-500 500] HU.

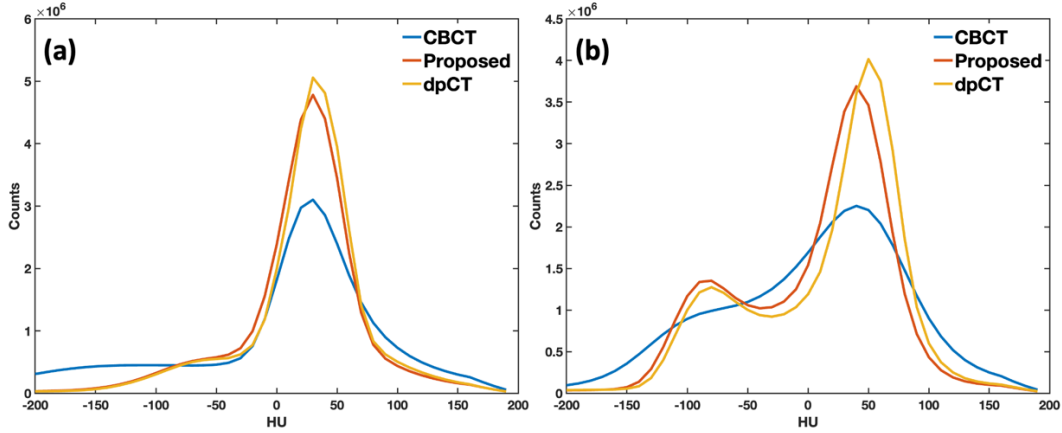

**Figure 6.** HU histogram plots of (a) brain and (b) H&N patient studies.

## 3.2. Quantitative analysis

Figures 7 shows the CBCT, generated sCT, dpCT images and corresponding error maps for brain and H&N studies. Consistent with artifact correction and HU fidelity improvement, the error distribution of sCT from dpCT is sparser compared to the difference map between CBCT and dpCT. In the brain patient study, our method improved the HU accuracy from MAE of 40.63 HU for CBCT to 25.99 HU for sCT. PSNR and NCC were 30.49 dB and 0.99 for sCT, compared to the values of 27.87 dB and 0.98 for CBCT. For the H&N patient study, MAE decreased to 32.56 HU from 38.99 HU while PSNR and NCC were slightly improved. Table IV tabulates the values of evaluation criteria for CBCT and sCT generated by the proposed method.

**Table IV.** Numerical comparison among CBCT and sCT generated by different diffusion models. Bold text indicates the best value in each metric.

|  | MAE (HU) |  | PSNR (dB) |  | NCC |  |
|---|---|---|---|---|---|---|
|  | Brain | H&N | Brain | H&N | Brain | H&N |
| CBCT | 40.63±12.71 | 38.99±14.07 | 27.87±2.20 | 27.00±1.98 | 0.98±0.01 | 0.98±0.01 |
| Uncond-1 | 61.98±21.13 | 54.83±21.11 | 23.29±2.22 | 24.45±2.37 | 0.95±0.02 | 0.96±0.01 |
| Uncond-2 | 34.68±12.66 | 36.32±12.57 | 28.07±2.37 | 27.12±1.95 | 0.98±0.01 | 0.98±0.01 |
| Adap-Cond-1 | 49.39±20.98 | 45.77±14.04 | 24.36±2.67 | 24.87±1.73 | 0.96±0.02 | 0.96±0.01 |
| Adap-Cond-2 | 84.15±31.44 | 82.89±25.19 | 20.27±2.07 | 20.78±1.69 | 0.91±0.04 | 0.91±0.02 |
| **Proposed** | **25.99±11.84** | **32.56±12.86** | **30.49±3.73** | **27.65±2.41** | **0.99±0.01** | **0.98±0.01** |

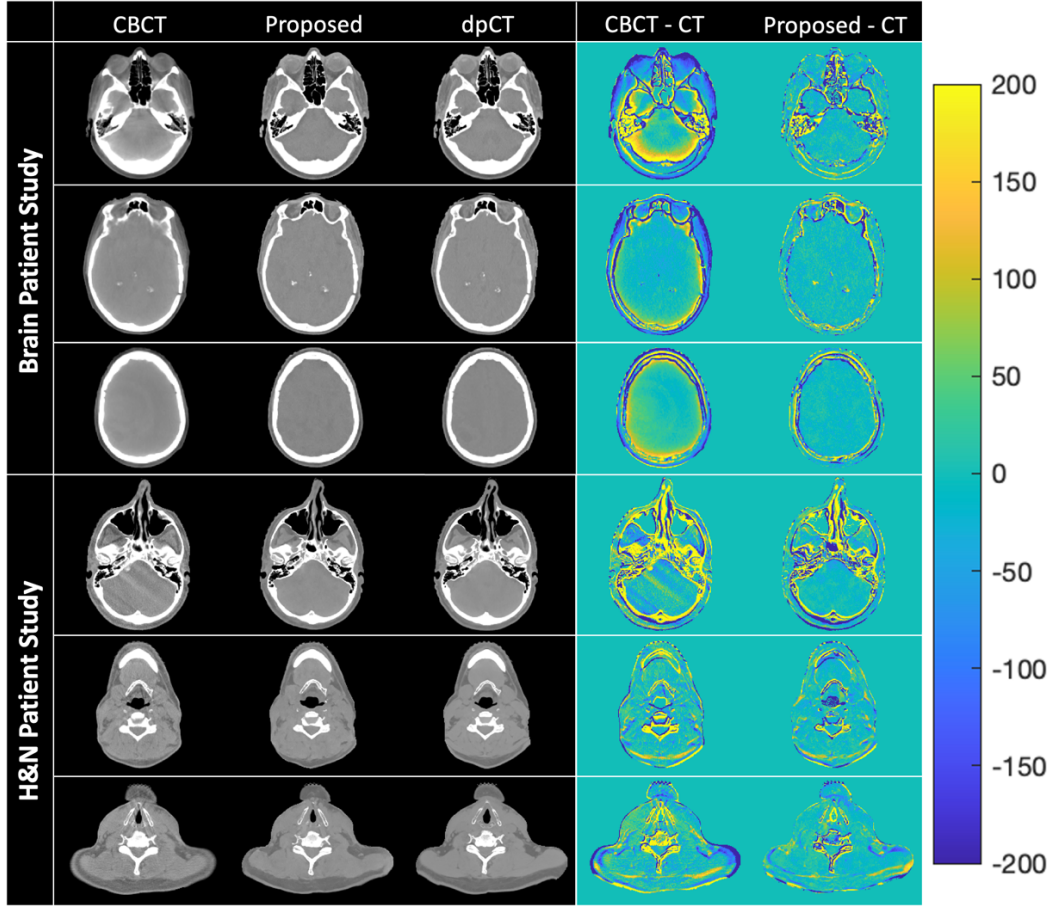

**Figure 7**. Summary of CBCT, sCT, dpCT images and difference maps for the brain and H&N patient studies. The display windows are [-500 500] HU for CT images and [-200 200] HU for error maps.

### 3.3. Comparison studies

sCT generated by different conditional schemes of DDPM are summarized in figure 8. For the Uncond-1 scheme, known as ILVR method, the generated sCT kept a similar outline with input CBCT while structural distortions and artifacts can be observed. This result is explainable because the ILVR only tries to preserve the coarse structure of CBCT by high-level semantics sharing. The Uncond-2 strategy performed better than ILVR because the image condition was directly enforced during the sampling process and more details were preserved on the generated sCT images. However, residual artifacts can remain if the original artifacts are severe because reweighted corrupted CBCT were added to the inferred image at each time step. For the two conditionally trained models with weak adaptive condition, Adap-Cond-1 method preserved most anatomy while few distorted structures can be observed, while in contrast Adap-Cond-2 scheme cannot even preserve the body contour in some cases because the guidance is too weak at the early stages of sample generation.

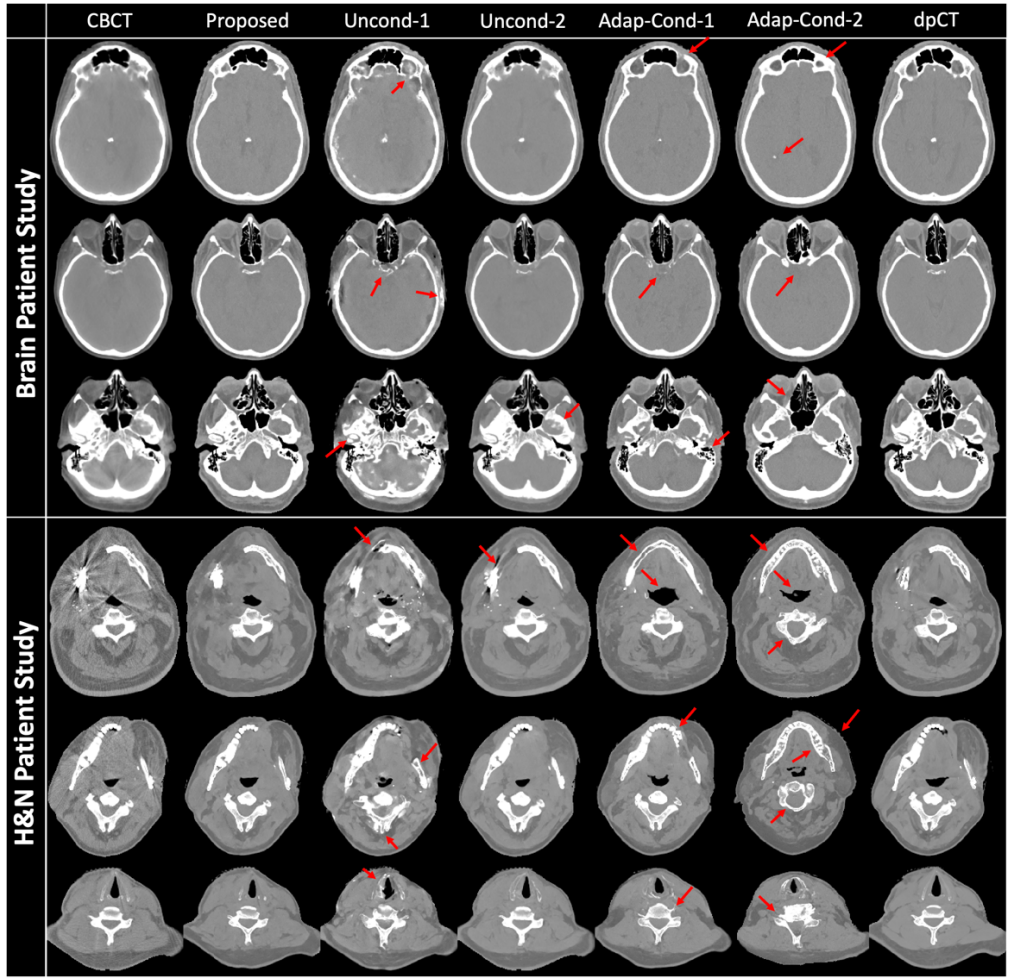

**Figure 8**. Summary of sCT generated by different diffusion models. Red arrows point out the structural distortions or the residual artifacts on sCT produced by four compared methods.

## 4. Discussion

This work formulated the CBCT-based CT synthesis task in a conditional DDPM framework. To the best of our knowledge, this is the first reported work to perform CBCT-CT translation using a diffusion model. Compared to previously proposed one-step domain transforming models like U-net and GAN, the diffusion model gradually translates a sample from one domain to another through a chain of Markov processes.

It has been shown in previously reported works that diffusion models outperform other generative models like GAN in the image generation, including the medical imaging tasks.[26,38] Thus, in this work, we focused on comparing the proposed method with other diffusion models rather than other generative models.

One of the common issues of supervised-learning schemes for medical image translation is the lack of exactly matched image pairs required for training. For example, it is impossible to obtain the matched CBCT-CT image pairs from one patient with and without metal artifacts. However, in the H&N patient study, the proposed conditional DDPM method was able to generate sCT samples with significantly reduced metal artifacts, which are superior to the artifact corrupted dpCT counterparts used for training. It shows that conditional diffusion models may be another solution for image translation tasks with inexact image pairs in addition to the unsupervised learning schemes. In this sense, the quantitative metrics that comparing sCT with dpCT are unfair to some extent because the referenced dpCT is not exactly matched with sCT in anatomy and sCT seems closer to an artifact-free ground truth.

It is also worth further investigating the feasibility of unsupervised diffusion models in sCT generation. It is important to note that unsupervised learning does not mean unconditional generative models. The first terminology refers to the fact that no paired data is needed for the model training, while the latter one indicates no control of the image generation. For medical image translation tasks like this study, the generated sCT is expected to share the same structure as the CBCT, which is exactly the condition in sCT generation. According to the principle of diffusion model, the only way to realize unsupervised conditional diffusion model is to perform unconditional training and conditional sampling. In this way, the condition enforcement in the standard form of conditional probability is no longer accessible. We will investigate possible solutions to this problem in future work.

In principle, diffusion models perform multi-step domain transform by predicting the noise distribution in each step. The stochastic property of the Gaussian noise makes it hard to incorporate additional prior information into the training process like the regularization strategy in other one-step image translation works. It means that the sampling quality relies only on the manifold-learning ability of networks without any guidance from prior information, which is insufficient to generate samples with inter-channel correlations like the dual-energy CT (DECT) synthesis task. To further improve the quality of generated samples, we will explore the possibilities of enforcing additional constraints or regularizations on the diffusion and reverse processes and investigate its feasibility on the synthetic DECT generation problem.

The time-consuming sampling procedure is the major obstacle to clinical practice of diffusion model-based methods while the training time is comparable to other one-step models like U-net. For a trained diffusion model with $T$ time steps, the generator will be implemented for $T$ times, resulting in a total sample generation time that is approximate $T$ times that of U-net or GAN models, making it impractical for online ART implementation. Accelerated sampling algorithms for diffusion models are a current hot research topic, and various strategies based on knowledge distillation[41,42] and training-free samplers have been proposed.[43,44] Previously reported work has evaluated the performances of different acceleration strategies[35] for the low-dose CT denoising problem and we will further investigate and determine the optimal algorithm for the medical image translation scenario in our future works.

The proposed method is based on 2D slice without a patch extraction strategy. It is reasonable to expect that the 3D patch-based models could produce samples with better structural preservation and better capture of spatial relationships at additional dimensions. Finally, diffusion models are still a developing field and more alternatives to DDPM will likely appear, which may outperform DDPM in the sCT generation problem.

## 5. Conclusions

In this work, we developed a conditional DDPM framework to perform sCT generation from CBCT images. The produced sCT images showed improved HU accuracy which make them suitable for the use of dose calculation. Much fewer artifacts were found in the sCT compared to input CBCT, which could give better results in CBCT-based online segmentation and replanning. The proposed method allows further quantitative applications of CBCT and enhance the potential of clinical practice of CBCT-based adaptive radiotherapy.

**Acknowledgement**

This research is supported in part by the National Institutes of Health under Award Number R01CA215718, R56EB033332, R01EB032680, and P30CA008748.